\documentclass[10pt,journal,compsoc]{IEEEtran}
\ifCLASSOPTIONcompsoc
  \usepackage[nocompress]{cite}
\else
  \usepackage{cite}
\fi
\usepackage{enumitem}
\usepackage{moreverb,url}
\usepackage[english]{babel}
\usepackage[colorlinks,bookmarksopen,bookmarksnumbered,citecolor=red,urlcolor=red]{hyperref}
\usepackage{breakurl}
\usepackage{color}
\usepackage{listings,multicol}
\usepackage{subfig}
\usepackage{array}
\usepackage{algorithm}
\usepackage{algpseudocode}
\usepackage{hhline}
\definecolor{AshGrey}{rgb}{0.7,0.75,0.71}
\definecolor{trolleygrey}{rgb}{0.5, 0.5, 0.5}
\definecolor{utahcrimson}{rgb}{0.83, 0.0, 0.25}
\definecolor{ao(english)}{rgb}{0.0, 0.5, 0.0}
\definecolor{cadmiumgreen}{rgb}{0.0, 0.42, 0.24}

\ifCLASSINFOpdf
  \usepackage[pdftex]{graphicx}
  \graphicspath{{images/}}
  \DeclareGraphicsExtensions{.pdf,.jpeg,.png}
\else
  \usepackage[dvips]{graphicx}
  \graphicspath{{images/}}
  \DeclareGraphicsExtensions{.eps}
\fi

\lstdefinestyle{base}{
  basicstyle=\scriptsize,
  language=C++,
  showspaces=false,
  showstringspaces=false,
  breakindent=3pt,
  breaklines,
  captionpos=b,
  breakatwhitespace=true,
  numbersep=5pt,
  keywordstyle=\color{blue},
  tabsize=2,
  moredelim=**[is][\color{red}]{@}{@},
  moredelim=**[is][\color{green}]{|}{|},
}

\newcommand{\V}{\verb}      
\newcommand{\TT}{\texttt}      

\newcolumntype{C}[1]{>{\centering\let\newline\\\arraybackslash\hspace{0pt}}m{#1}}
\newcolumntype{L}[1]{>{\raggedright\let\newline\\\arraybackslash\hspace{0pt}}m{#1}}
\newcolumntype{R}[1]{>{\raggedleft\let\newline\\\arraybackslash\hspace{0pt}}m{#1}}

\begin{document}

\title{CRAFT: A library for easier \\
application-level Checkpoint/Restart and \\
Automatic Fault Tolerance}

\author{Faisal~Shahzad,
        Jonas~Thies,
        Moritz~Kreutzer,
        Thomas~Zeiser,
        Georg~Hager,
        and~Gerhard~Wellein
\IEEEcompsocitemizethanks{\IEEEcompsocthanksitem  F. Shahzad, M. Kreutzer, T. Zeiser, 
G. Hager and G. Wellein are affiliated with Erlangen Regional Computing Center (RRZE),
University of Erlangen-Nuremberg, Martensstrasse 1, 91058, Erlangen, Germany.\protect\\
E-mail: faisal.shahzad@fau.de 
\IEEEcompsocthanksitem J. Thies is with the German Aerospace Center (DLR), 
Simulation and Software Technology, Linder H\"ohe, 51147, Cologne, Germany.}
}

\IEEEtitleabstractindextext{%
\begin{abstract}
In order to efficiently use the future generations of supercomputers, fault 
tolerance and power consumption are two of the prime challenges anticipated by the 
High Performance Computing\,(HPC) community.
Checkpoint/Restart\,(CR) has been and still is the most widely used 
technique to deal with hard failures. 
Application-level CR is the most effective CR technique
in terms of overhead efficiency but it takes a lot of implementation effort.

This work presents the implementation of our C++ based library CRAFT 
(Checkpoint-Restart and Automatic Fault Tolerance), which serves two purposes.
First, it provides an extendable library that significantly eases the 
implementation of application-level checkpointing.
The most basic and frequently used checkpoint data types are already part of CRAFT 
and can be directly used out of the box.
The library can be easily extended to add more data types.
As means of overhead reduction, the library offers a build-in asynchronous 
checkpointing mechanism and also supports the Scalable Checkpoint/Restart\,(SCR)
library for node level checkpointing.
Second, CRAFT provides an easier interface for User-Level Failure 
Mitigation\,(ULFM) based dynamic process recovery, which significantly reduces
the complexity and effort of failure detection and communication recovery 
mechanism.
By utilizing both functionalities together, applications can write 
application-level checkpoints and recover dynamically from process failures 
with very limited programming effort.

This work presents the design and use of our library in detail. 
The associated overheads are thoroughly analyzed using several benchmarks. 
\end{abstract}

\begin{IEEEkeywords}
Application-level checkpoint/restart, 
automatic fault tolerance, User-Level Failure Mitigation\,(ULFM)
\end{IEEEkeywords}}

\maketitle

\section{Introduction}
\label{sec:intro}
\IEEEPARstart{T}{he} ever increasing demand of more computational power continuously leads to 
the deployment of larger systems with more efficiency. 
After the halt in the increase of the processor frequency, the consistent 
growth in larger clusters is the result of the growing level of 
hardware parallelism. 
This results in a decrease of the mean time to failure (MTTF) with every
new generation of large clusters.
For example, the BlueGene/P system `Intrepid' (debuted at $\#$4 in the June 2008 
top500\footnote{Top500 Website: \url{http://top500.org} } list) had a 
reported mean time to (hardware) interrupt of 7.5 days 
\cite{addressing-failures-inexacale},
whereas the more recent BlueGene/Q system `Sequoia' (debuted at $\#$3 in the Nov. 
2013 top500 list) has a reported node failure rate of 1.25 per day 
\cite{Invited_talk_Jackdongara}.
This trend raises the concerns in the HPC community about the effective 
usability of clusters at the exascale level.

A program running on HPC systems can fail for many reasons, e.g., 
hardware and software faults, silent errors, Byzantine failures, etc. 
A study by El-Sayed et. al. \cite{El-Sayed:2013:RLF_Bianca} has found that 
60\% of all failures are attributed to either memory or CPU failures.
Such failures, in addition to many others, lead to process failure and eventually
to the failure of the MPI-job as a whole. 

There are many fault tolerance techniques to reduce the damage of 
faults to a minimum. 
These can be classified as one or a 
combination of the following four categories \cite{hursey-phd-thesis-2010}: 
algorithm-based fault tolerance\,(ABFT), checkpoint/restart(CR), 
message logging, and redundancy.
These categories vary by the faults they can detect and/or recover, 
their coverage, etc. The coverage of a fault tolerance technique is 
defined as the measure of its effectiveness 
\cite{Avizienis:2004:basic-concepts-taxonomy}.
None of the fault tolerance techniques can guarantee 100\% coverage 
of all possible failures, e.g., a technique against silent errors can not 
protect the application against hardware failures.
Due to its characteristic properties, CR
is the most widely used fault tolerance technique. 
The others, such as message logging and 
ABFT, also often use CR as a supporting component.
Apart from fault tolerance, many HPC applications use CR 
to cope with other issues such as maximum walltime limit, which is present on 
almost all large production systems.

In parallel applications, CR services are categorized on two different levels:
the mode of saving the process state, and the communication handling method during
checkpointing.

The communication is handled either in an uncoordinated or a coordinated 
approach.
In an uncoordinated checkpointing approach, each process takes its checkpoint 
independently without any notification to other processes. 
Thus the communication channels are not flushed and thereby are not in a consistent 
state.
Despite its simplicity, this is generally not a preferred method
due to its drawbacks in the restart phase and a large requirement of storage capacity
\cite{OnCoordinatedCheckpointing-Cao, AComparisonBetweenCPSystems}.
In contrast, a coordinated approach creates a checkpoint of all 
processes at logically the same time.
Depending on whether the communication channels are blocked or not during 
checkpointing, the coordinated checkpointing is further divided into blocking 
and non-blocking approaches 
\cite{hursey-phd-thesis-2010, BlockingVsNonBlockingCoordinatedCP-2006}.

The methods for saving the state of the processes are categorized as follows 
depending on the level of transparency and location of implementation 
in the software stack \cite{hursey-phd-thesis-2010}.
1) System-level: As its name suggests, this checkpoint is implemented on kernel
level. Thus, the whole memory footprint of the application is checkpointed. 
2) User-level: Implemented in the user-space, such a CR service captures the 
process state by virtualizing corresponding system calls to the kernel without being 
tied to the kernel itself.
3) Application-level: The user manually determines the data that 
needs to be checkpointed. This offers the possibility to 
save the minimum amount of data needed for a checkpoint and thus incurs minimum 
checkpoint overhead. 

The application-level CR\,(ALCR) approach, though attractive 
\cite{system-level-vs-user-defined-CP}, carries many inherent challenges.
In order to save the minimum essential data, the user must be able 
to distinguish the data that is necessary to checkpoint.
We define a `checkpoint' as the collection of all data objects that are necessary 
to recover a particular stage of the program. 
A program can have multiple/nested checkpoints at various stages, 
each having a different checkpoint interval.
The updated versions of the same checkpoints are defined as `checkpoint-versions' 
(CP-version). 
Each checkpoint can contain various data types, such
as plain-old-data\,(POD), POD-arrays, POD-multiarrays, and even complex 
user-defined class objects. 
Moreover, any data object in a program can have parts which
are not necessary for checkpointing. 
Secondly, the user must write the functions to save that data in a secure 
location (e.g., a parallel file system\,(PFS)).

The first part of this work focuses on simplifying the method of including
ALCR in MPI applications. 
We do this by implementing the CR relevant functions (read, write, etc.) for different 
data types and provide an interface for them. 
We say that a data types is `CRAFT-checkpointable', if its CR-related functions 
are linked to CRAFT.
The application developer can add ALCR with a minimal effort and changes in the application.
The most common, standard and frequently used data types 
(e.g. POD, POD-arrays, POD-multiarrays etc.) are contained in the default 
CRAFT-supported checkpointable data types and can be used directly out of the box.
In addition, the user can make any arbitrary data type CRAFT-checkpointable
by a simple extension mechanism, which essentially requires to implement its 
CR-related functions.
The CRAFT library does not entail any communication handling protocol for CR 
phases, thus it is the user's responsibility to insert the CR routines in a 
consistent communication phase.

\begin{figure}[tb]
	\centering
	\includegraphics[width=5cm,clip=true]{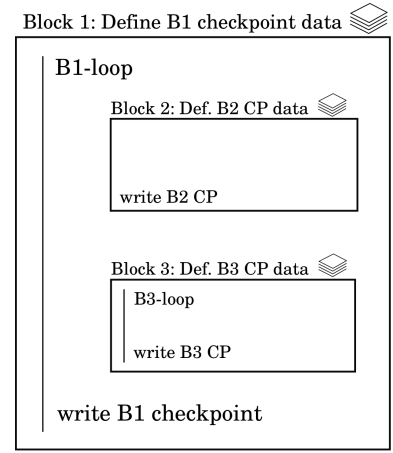}
	\caption[]{A sketch of a potential CRAFT application. 
The program consists of blocks where the checkpoint data of each block is defined
at the start of the block. 
Each block may subsequently contain more sub-blocks and/or loop bodies. 
The checkpoints are written at the end of a block/loop body.
	\label{fig:craftProgramBlockDiag}}\vspace{.00cm}
\end{figure}

This CR approach is widely applicable as long as the application flow does not 
contain unpredicted branches with sudden jump/break points.
A sample control flow of a program that can use CRAFT's CR solution is shown in Fig.
\ref{fig:craftProgramBlockDiag}.
The program can be seen as combination of execution blocks,
where a block may contain further blocks or a loop body. 
In order to create a checkpoint inside a loop, the data items which it entails, 
must be completely defined before entering the loop body.
The checkpoints are written at the end of the block/loop body.
This structure encompasses a large variety of applications, e.g.,
explicit time integration schemes for solving Partial Differential 
Equations\,(PDE) (1 loop), implicit time stepping with a Newton-Krylov non-linear 
solver (3 nested loops), continuation and linear stability analysis by
solving a nonlinear PDE for a sequence of parameters and solving an eigenvalue 
problem at each steady state (a loop with two blocks inside, each containing 
a nested loop), etc.

The frequency of the checkpoints (interval) is a critical parameter in this 
context, and is well-studied in the literature. 
A model for the optimal checkpoint interval can be found in 
\cite{Daly:a-model-for-predicting-cp-time}.

Even in the presence of CR functionality in an application, 
the fail-stop failures lead to process/node failures(s),  
which eventually cause job abortion.
Larger jobs have a much higher probability to encounter such failures, 
which means spending additional time in the queuing process of a typical cluster and 
then restarting the job.
Thus it is necessary to develop ways in which applications can automatically 
recover from process failure(s), which we define here as 
`Automatic Fault Tolerance'\,(AFT). 
The second part of the library deals with AFT. 
Among other aspects of developing an AFT-application, 
the communication recovery is the most crucial one. 
For this purpose, we use User-Level Failure Mitigation\,(ULFM) 
\cite{bland2012proposal}, which is a recently developed prototype implementation 
of a fault tolerant MPI, and aims to be included in the MPI-4.0 Standard in 
the near future.
The ULFM provides MPI-extensions by which the user can detect, 
acknowledge, and eventually recover from a process failure in a communicator.
The ULFM extensions provide the fault tolerant MPI functions but
leave the actual recovery method implementation to the user. 
Thus, the user can be flexible about the communication recovery process, e.g.,
shrinking recovery, non-shrinking recovery etc. 
In CRAFT, we have fixed the fault detection method to exception handling, 
and thereby hidden away many fault detection and communication recovery details 
behind a very simple-to-use API. 

The main contributions of our work are summarized as follows:
\begin{enumerate}[leftmargin=1\parindent]
    \item Design and implementation of a C++ fault tolerance library, CRAFT, that 
          provides an easy interface for making application-level checkpoints 
          for a variety of data types. 
          In addition, the user can extend the library for any arbitrary data type.

    \item The extendability of CRAFT is shown by creating a simple class 
          and showing the steps to make it a CRAFT checkpointable 
          data type.

    \item A built-in asynchronous checkpointing mechanism can 
          be used to reduce the checkpointing overhead. In addition, CRAFT also supports 
          the Scalable Checkpoint Restart\,(SCR) library, 
          which enables the checkpoint storage and recovery at the node level.

    \item An easier interface is provided to embed Automatic Fault 
          Tolerance in applications so that they can dynamically recover 
          from process failure(s). 
          The user can choose between two different communication recovery models:
          shrinking or non-shrinking.

    \item A scaling analysis of the automatic failure recovery (AFT) feature of the 
          library is performed on up to 2560 processes on 128 nodes. 
          In addition, a Lanczos eigenvalue solver is used to showcase and analyze the 
          overheads of its CR as well as AFT features combined.

\end{enumerate}

The rest of the paper is organized as follows.
The design, implementation, and interface details of the CRAFT's CR functionality 
are presented in Section \ref{sec:craftCR}. 
Section \ref{sec:craftAFT} presents the design and implementation of AFT in CRAFT.
Generic properties of CRAFT are presented in Section \ref{sec:craftProperties}.
The details of benchmark applications, software environment and testbed systems
are described in Section \ref{sec:experimental_framework}.
The scalability as well as the overheads involved due to CR and AFT features of CRAFT 
are analyzed in detail in Section \ref{sec:results}. 
In Section \ref{sec:relatedWork}, a brief summary of the related work is presented.
Section \ref{sec:summary} presents a summary and concludes the paper.

The CRAFT library is open source under a BSD license and is available at 
\cite{website:craftlib}.

\section{CRAFT(I): Checkpoint/Restart library}
\label{sec:craftCR}

This section presents the implementation details of the application-level 
CR functionality of CRAFT.

\begin{table*}[bt]
\begin{tabular}{@{}p{\dimexpr0.5\linewidth-\tabcolsep}
                  p{\dimexpr0.5\linewidth-\tabcolsep}@{}}
\begin{lstlisting}[style=base, caption=A sample application without application-level checkpoints., label=algo:sampleApp1]
#include <mpi.h>  

int main(int argc, char* argv[]){
	int n=5, iteration=1;
	double dbl = 0.0;
	int * dataArr 	= new int[n];







  for(; iteration <= 100 ; iteration++){
    // Computation-communication loop
    modifyData(&dbl, dataArr);

	}
	return EXIT_SUCCESS;
}
\end{lstlisting}
&
\begin{lstlisting}[style=base, caption= Corresponding changes in Listing \ref{algo:sampleApp1} for CRAFT enabled application-level checkpoints., label=algo:sampleApp1WCR]
#include <mpi.h>  
#include <craft.h>
int main(int argc, char* argv[]){
	int n=5, iteration=1, cpFreq=10;
	double dbl = 0.0;
	int * dataArr 	= new int[n];
  // ===== DEFINE CHECKPOINT ===== // 	
	@Checkpoint  myCP("myCP", MPI_COMM_WORLD);@
	@myCP.add("dbl", &dbl);@
	@myCP.add("iteration", &iteration);@
	@myCP.add("dataArr", dataArr, n);@
	@myCP.commit();@
  @myCP.restartIfNeeded(&iteration)@;
  for(; iteration <= 100 ; iteration++){
    // Computation-communication loop
    modifyData(&dbl, dataArr);
    @myCP.updateAndWrite(iteration, cpFreq);@
	}
	return EXIT_SUCCESS;
}
\end{lstlisting}
\end{tabular}
\label{table:toycode-craft}
\end{table*}


We start by looking at an example in which checkpoints are 
created using CRAFT. 
Listings \ref{algo:sampleApp1} and \ref{algo:sampleApp1WCR}
draw a comparison between a simple 
iterative toy application and its corresponding CRAFT-enabled 
version, respectively.
A \V|Checkpoint| object is created and all relevant data is added to it
using the \V|add()| member function.
In this example, a double element \V|dbl|, the iteration counter 
\V|iteration| and an integer array \V|dataArr| of length \V|n| 
are added.
A pointer to each checkpointable object is saved. 
If the user has enabled the asynchronous checkpointing option, a copy of the 
added checkpointable object is created at this stage.
Once all relevant data was added, the \V|Checkpoint| object is committed, 
which means that no further data can be added.
The application uses the original data for computation/communication 
in the usual manner.
The \V|updateAndWrite()| member function updates and writes all checkpoint 
data at the iterations that match with the checkpoint frequency.
The member function \V|restartIfNeeded()| determines if the program is restarted,
in which case it reads the checkpoint. 

\subsection{Design}
\begin{figure*}[htb]
	\centering
	\includegraphics[width=14.5cm,clip=true]{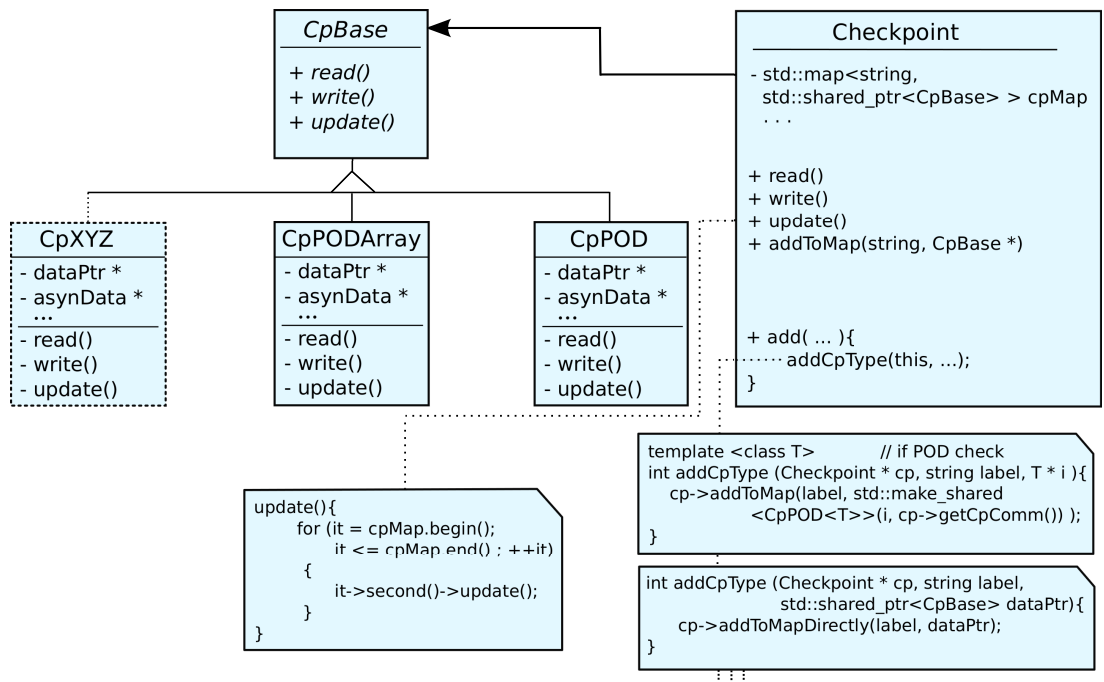}
	\caption[]{Design logic of the CRAFT library. 
The user creates a checkpoint by the adding checkpointable objects via the 
\texttt{Checkpoint::add()} member function, which eventually puts these objects in 
\TT{cpMap}. 
The \texttt{Checkpoint::read()}, \texttt{write()} and \texttt{update()} methods 
iterate thought the \TT{cpMap} and call each object's corresponding calls.
Each checkpointable data type class must be inherited from \TT{CpBase} and must 
contain the implementation of its CP-related virtual functions (i.e., \texttt{read()}, 
\texttt{write()}, and \texttt{update()}).
	\label{fig:craftDesignLogic}}\vspace{.00cm}
\end{figure*}

The design logic of the CR feature of the CRAFT library is explained in Fig. 
\ref{fig:craftDesignLogic}. 
The core part consists of a base class (\V|CpBase|) with three pure virtual 
functions called \V|read()|, \V|write()|, and \V|update()|. 
For each checkpointable data type, a class is derived from \V|CpBase| 
which carries the implementation of its virtual functions.
The user interacts with CRAFT via a \V|Checkpoint| class, which contains a C++ 
standard map (\V|cpMap|) that gathers the shared pointer (\V|std::shared_ptr|) of all 
checkpointable data types that are added by the user in a specific checkpoint. 
The \V|Checkpoint| class also contains the interface for adding the checkpointable objects
from the user application. 
In the application, the user defines a checkpoint by defining an object of the
\V|Checkpoint| class and adding all necessary checkpointable objects into it as
described in the subsection above.

\subsection{CR interface}
The most important functions and data members of the CRAFT CR interface are 
listed below along with a brief description.\\

\texttt{Checkpoint::Checkpoint(string cpName, MPI\_Comm comm)}:
The constructor of a \V|Checkpoint| object takes two inputs. 
The first is a unique name, which is used to create each checkpoint specific 
directory within the base checkpoint-path, thus the checkpoint name has to 
be a valid directory name. 
The base checkpoint-path can be provided by setting CRAFT environment variables (see 
Sec. \ref{sec:craft-params}).
The second argument (optional) can be used to specify the communicator 
attached to the \V|Checkpoint| object. 
For using the AFT feature of CRAFT, a separate communicator other than 
\V|MPI_COMM_WORLD| must be provided (see Sec. \ref{sec:craftAFT} for details). 

\texttt{Checkpoint::add(string key, ... )}:
This member function adds the checkpointable data objects 
to the \V|Checkpoint| object.
The first \V|std::string key| argument is used to create the file name of 
each added object of a particular checkpoint, thus it has to be a valid 
file name. 
If the user creates a checkpoint on a PFS using MPI-IO, the 
rank-independent file names are generated. However, if the user does not use MPI-IO, 
or creates node level checkpoints by enabling SCR, process-local file names 
are generated containing the rank of the process.
These full file names are passed to each checkpointable's \V|read()| and 
\V|write()| member functions.
The \V|add()| member function has multiple overloaded implementations 
depending on the type of data that needs to be checkpointed.
By default, checkpoints for following data types can be added.
\begin{itemize}[leftmargin=*] 
\item POD: \texttt{add(string key,POD* dat)} can be used to add a POD element where 
        \TT{dat} could be a pointer to an \TT{int}, \TT{double}, \TT{float}, or a \TT{std::complex} element.
\item POD Array: \texttt{add(string key,POD* dat,int n)} can be used to add an array of 
        POD elements, where \TT{n} is the length of the array.
\item POD Multi-array: \texttt{add(string key, POD** dat, int n, int m, int toCpCol)}
        can be used to add a 2D POD array of size \TT{n}$\times$\TT{m}. 
        The user can further specify a specific column that needs 
        checkpointing out of a POD multi-array.
\item MPI data type: \texttt{add(string key,void * buff, MPI\_Datatype * data)} adds
        an object of an MPI derived data type.
        For asynchronous writes, CRAFT uses \V|MPI_Pack()| to copy the data in 
        a separate buffer, which is then saved asynchronously.
\item \TT{CpBase} derived types: \TT{add(string key, CpBase * data)} can be used to 
        add objects of data types which were derived from \TT{CpBase} by the user.
\end{itemize}

Furthermore the user can extend the library and add any arbitrary data type by implementing 
a specialized function (discussed in detail in Section \ref{sec:craft-extension}).

\texttt{Checkpoint::commit()}: Once all the checkpoint relevant data is added to 
the \V|Checkpoint| object, it is committed, i.e., no data can be further added in 
later stages of the program to the same \V|Checkpoint| object. 
The user may create a new instance of the \V|Checkpoint| class at a different stage of 
the program to checkpoint different data.

\texttt{Checkpoint::updateAndWrite()}: This method iterates on the \V|cpMap|
and calls the corresponding \V|update()| (in case of asynchronous checkpointing) 
and \V|write()| member function on each of the objects in the container. 
The method (optionally) can take two input arguments \V|(int iteration , int cpFreq)|, 
where a checkpoint is only created at every \V|cpFreq| of the \V|iteration| counter 
(i.e., \V|if(iteration 

\texttt{Checkpoint::restartIfNeeded()}: This member function checks if this application run is
a restarted run by looking for any previous checkpoints.
If there are any, it also reads the data.
This behavior can be overridden by setting the \V|CRAFT_READ_CP_ON_RESTART|
environment variable appropriately (see Sec. \ref{sec:craft-params}), 
through which the program can be re-executed without reading previous checkpoints.

\texttt{Checkpoint::wait()}: The user can choose to make an asynchronous checkpoint 
without generating an additional copy of the data 
(see Sec. \ref{sec:craft-params}).
In this case, \V|wait()| should be used to make sure that 
asynchronous writing of the checkpoint has finished before modifying the data.

\subsection{CRAFT extension}
\label{sec:craft-extension}
In order to make an arbitrary new data type CRAFT-checkpointable, the user may 
follow any of the following three methods. 
These methods differ based on the access rights to the corresponding data type class.

\begin{lstlisting}[float=tb,style=base,caption={A sample class \texttt{rectDomain}, whose objects a user may want to checkpoint in the application.},label=algo:rectDomainClass]
class rectDomain{
public:
  rectDomain( const int length, const int width);
  rectDomain( const rectDomain &obj );
	~rectDomain();
private:
	int length; 
	int width;
  double * val; 
};
\end{lstlisting}

\begin{lstlisting}[float=tb,style=base,caption={Assuming the \texttt{rectDomain} class is unmodifiable, a wrapper class \texttt{cpRectDomain} is created that implements the CP-related (\texttt{read()}, \texttt{write()}, and \texttt{update()}) functions.},label=algo:cpRectDomainClass]
#include "cpBase.hpp"
class cpRectDomain: public CpBase
{
public:
  cpRectDomain( rectDomain * dataPtr_, const MPI_Comm cpMpiComm_=MPI_COMM_WORLD){
    dataPtr = dataPtr_;
    *asynData = *dataPtr_;
  }
	~cpRectDomain(){}	
private:
	rectDomain * dataPtr;
	rectDomain * asynData;
  int update(){
    // update asynData from dataPtr
  }
	int write(const std::string * filename){
    // write asynData to the given filename
  }
	int read(const std::string * filename){
    // read asynData to the given filename
  }
};
\end{lstlisting}

\begin{lstlisting}[float,style=base,caption={An Example for CRAFT-based checkpointing of objects of the \texttt{rectDomain} data type.},label=algo:rectDomainApp]
#include <craft.h>
#include <rectDomain.h>
...
int main(int argc, char* argv[]){
  ...
	int iteration = 1, n = 100, cpFreq = 10;
  @rectDomain myRecDom(3, 4);@

	Checkpoint myCP( "myCP", MPI_COMM_WORLD);
	myCP.add("iteration", &iteration);
  @myCP.add("myRecDom", std::make_shared<cpRectDomain>(myRecDom));@
	myCP.commit(); 
  myCP.restartIfNeeded(&iteration);
  for(; iteration <= n; iteration++){
    // Computation-communication loop
    myCP.updateAndWrite(iteration, cpFreq);  
	}
  ...
}
\end{lstlisting}

\begin{lstlisting}[float=tb,style=base,caption={The interface function between \texttt{cpRectDomain} data type and the \texttt{Checkpoint::add()} member function.},label=algo:addedCpTypesRectDomain]
#include "cpTypes/cpRectDom/cpRectDom.hpp"
int addCpType(Checkpoint * cp, std::string label, rectDomain * dataPtr){
  cp->addToMap(label, new cpRectDomain(dataPtr, cp->getCpComm()));
}
\end{lstlisting}

\begin{enumerate}[leftmargin=1\parindent]
\item If the user can modify the target data type class, he can simply 
inherit CRAFT's \V|CpBase| class and implement its required CP-related
virtual functions. 
In this way, the original class itself becomes CRAFT-checkpointable 
and the \TT{Checkpoint::add()} member function can be directly used on its objects
in the application.

\item For one or the other reason, if the user can not (or does not want to) 
modify the target class, a checkpointable wrapper class can be made by 
inheriting CRAFT's \V|CpBase| class. 
The data type's CP-related virtual functions must now be implemented 
inside this wrapper class. 
The \TT{Checkpoint::add()} method can now take the checkpointable wrapper objects 
as its arguments.
An example of this extension methodology is shown in Listings \ref{algo:rectDomainClass}, 
\ref{algo:cpRectDomainClass}, and \ref{algo:rectDomainApp}. 
Listing \ref{algo:rectDomainClass} shows an example class,
\TT{rectDomain}, whose objects the user may like to checkpoint in the application.
Assuming that the \TT{rectDomain} class itself cannot be modified, a wrapper class 
(\TT{cpRectDomain}) is created as shown in Listing \ref{algo:cpRectDomainClass}.
Apart from the implementation of the required virtual functions of the \V|CpBase| 
class, the user can create an extra copy of the data for asynchronous checkpointing.
In the application, the user can now create this wrapper object on top 
of the original checkpoint object and use the \TT{add()} method as shown in
Listing \ref{algo:rectDomainApp}.
He may optionally add an interface function
inside CRAFT which creates the wrapper checkpointable class objects and 
adds them to the checkpoint map.
This option is interesting for library developers, who cannot (or wish not to) 
modify the original data type, but still want to offer 
the end user a possibility of using \TT{Checkpoint::add()} directly on the original 
data type objects without the hassle of creating wrapper class objects.
For the above mentioned \TT{rectDomain} example of Listing 
\ref{algo:rectDomainClass}, this interface function could look as shown in 
Listing \ref{algo:addedCpTypesRectDomain}.

\item In a similar approach as above, the user can also create the checkpointable 
wrapper class (similar to Listing \ref{algo:cpRectDomainClass}) by inheriting 
CRAFT's \V|CpBase| as well as the original \TT{rectDomain} class. 
This way the same objects can be used for normal computation/communication 
as well as for checkpointing purposes.
A more refined solution would be to use different namespaces, each having the 
same class names but one being a CRAFT-checkpointable class in addition.

\end{enumerate}

\subsection{CR optimizations}
\label{sec:cr-optimizations}
The CRAFT library supports asynchronous checkpointing 
\cite{Shahzad:2012:ACD:async-ckpt} 
and the Scalable Checkpoint Restart (SCR) library 
\cite{Moody:2010:multi-level-checkpointing-system} as means to reduce the CR 
overhead.

The asynchronous checkpointing can be controlled by using the 
\V|CRAFT_WRITE_ASYNC| environment variable.
CRAFT uses \V|std::future| and \V|std::async| to assign the asynchronous 
writing of checkpoints to a dedicated thread of each process.
This method of checkpointing, in its n\"aive implementation, requires
a separate copy of the data, thus writing and computation could 
take place simultaneously.
For this purpose, all checkpointable default data types in CRAFT 
create an additional copy, if this option is enabled.
The copy of the added data is updated from the original data in the 
\V|Checkpoint::update()| member function.
In case of extending CRAFT for a new data type, the user is responsible
for creating a copy along with its \V|update()| method to benefit from 
this feature. 
Alternatively, the user can opt to perform the asynchronous writes without
creating an additional data-copy. This can be done via the
\V|CRAFT_WRITE_ASYNC_ZERO_COPY| environment variable.
The user can then use \V|Checkpoint::wait()| to ensure the completion of 
the asynchronous write before modifying the data.
As the IO routines can potentially use MPI routines, enabling this feature 
requires full threading support (i.e., \V|MPI_THREAD_MULTIPLE|) from the underlying 
MPI library.
By setting the appropriate \V|CRAFT_ASYNC_THREAD_PIN_CPULIST| 
parameter, the asynchronous thread locality (pinning) 
information can be provided, thereby maximizing the performance gain of asynchronous threads. 
(see Sec. \ref{sec:craft-params} for details).

The SCR library provides the user with the option for node level
checkpointing. This can take one of three forms:
local node level, partner level, or partner-XOR level. 
In case of one node failure, partner level checkpointing allows to recover
restart data from the failed node's neighbor.
Apart from node level checkpoints, the less frequent 
checkpoints on PFS level can be made in order to enable recovery 
from multi-node failures.
For more information on SCR, 
see \cite{Moody:2010:multi-level-checkpointing-system}.
All necessary SCR-relevant changes, e.g., setup (\V|SCR_Init()|,
\V|SCR_Finalize()|) and checkpoint calls (\V|SCR_Start_checkpoint()|, 
\V|SCR_Route_file()|, and \V|SCR_Complete_checkpoint()|) are integrated into CRAFT.
Once CRAFT is compiled with SCR, the node level checkpoints 
are automatically enabled for CRAFT-checkpointable data without any modification 
in the user application.
The user can, however, disable SCR usage either partly (i.e., for specific 
checkpointable data using the \V|Checkpoint::disableSCR()| method) or completely 
(by setting the \V|CRAFT_USE_SCR| environment variable).
As our benchmark platform operates on a Torque resource manger, 
minor add-ons in SCR were required.
Furthermore, the usage of the AFT feature (Sec. \ref{sec:craftAFT}) in 
combination with SCR requires the SCR initialization with a fault tolerant communicator.

\begin{lstlisting}[float,style=base,caption={A pseudo-code example of nested checkpoints using CRAFT. The usage of SCR is disabled for CL1. Thus only high-freqency, smaller checkpoints of CL2 are stored on node-levels.},label=algo:nestedCp]
Checkpoint CL1("CL1", FT_Comm);
Checkpoint CL2("CL2", FT_Comm);
@CL2.subCP(&CP1);@
[...]   // add L1data
[...]   // add L2data
CL1.disableSCR();
|||**********************************************|||
CL1.restartIfNeeded()                        |||*|||
for( L1iter ---> nL1iter  )                  |||*|||
{                                            |||*|||
	/* L1 COMPUTATION COMMUNICATION */         |||*|||
  @****************************************@   |||*|||
	CL2.restartIfNeeded()                  @*@   |||*|||
	for( L2iter ---> nL2iter  ) {          @*@   |||CL1|||
 /* L2 COMPUTATION COMMUNICATION */      @CL2@ |||*|||
		CL2.updateAndWrite(L2iter, L2cpFreq);@*@   |||*|||
	}                                      @*@   |||*|||
  @****************************************@   |||*|||
	/* L1 COMPUTATION COMMUNICATION */         |||*|||
  CL1.updateAndWrite(L1iter, L1cpFreq);      |||*|||
}                                            |||*||| 
|||**********************************************|||
\end{lstlisting}

\subsection{Multi-level/nested checkpoints example}
\label{sec:multilayer-feature}
The CRAFT library can be used to create multi-level and nested checkpoints.
Multi-level checkpoints deal with \V|Checkpoint| objects defined in different 
successive stages of a program.
Nested checkpoints, on the other hand, are needed to checkpoint data  
in nested blocks of the program.
The latter can, for example, arise in the applications 
where a time-consuming outer loop encloses another time-consuming inner loop.
The results of the outer loop are mostly calculated based on the inner loop 
results.
In such cases, creating a single combined checkpoint at the inner level would lead 
to saving the redundant/unmodified data of the outer level.
Nested checkpoints require careful attention from the user;
e.g., all nested checkpoints must be committed before entering the first level.
Listing \ref{algo:nestedCp} shows a pseudo-code, where the CL2-checkpoint of 
level-2 (L2) is nested into the CL1-checkpoint of level-1 (L1).

\begin{figure}[htb]
  \centering
  \includegraphics[width=7.5cm,clip=true]{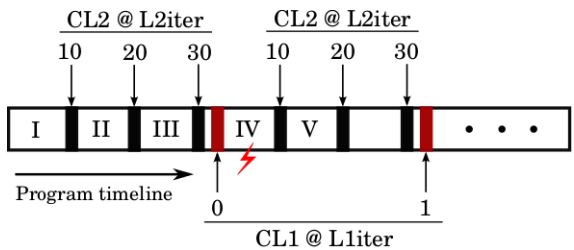}
	\caption[]{Program execution and checkpoint stages of the nested 
checkpoints of Listing \ref{algo:nestedCp}. Restarting from the latest
versions of all nested checkpoints can lead to data inconsistency.
	\label{fig:nestedCpTimeline}}\vspace{.00cm}
\end{figure}

A restart from a nested checkpoint requires careful consideration, 
as restarting all nested levels from their latest generated checkpoints
may lead to inconsistencies.
This problem is explained in Figure \ref{fig:nestedCpTimeline} and Table 
\ref{tab:nestedCpRecoveryStages}.
Figure \ref{fig:nestedCpTimeline} shows the stages of checkpoints 
corresponding to the nested checkpoint levels of 
Listing \ref{algo:nestedCp}.
The program's timeline follows the following parameters corresponding
to Listing 
\ref{algo:nestedCp}: \V|nL1iter=2|, \V|L1cpFreq=1|, \V|nL2iter=30|, 
\V|L2cpFreq=10|.
Table \ref{tab:nestedCpRecoveryStages} shows the correctly restarted 
stages of the program depending on the instance of failure occurrence. 
In case of a failure at stage `I', no checkpoint is read since none was created 
yet.
At stages `II' and `III', the latest CL2 checkpoint is read, whereas no 
CL1 checkpoint is read (i.e. CL2-v=10,20).
However, if the failure occurs at stage `IV', only the CL1 checkpoint 
should be read. 
Reading the latest CL2 checkpoint of CL2-v=30 will lead to inconsistency 
since the L2 iterations re-start from 0 at this stage. 
The CL2-v=30 checkpoint should be read only if failure happens after creating
this checkpoint and before writing CL1-v=0.
A failure at stage `V' restarts from CL1-v=1, CL2-v=10.

In CRAFT, we solve this inconsistency problem by invalidating the 
child checkpoints as soon as the parent checkpoints are fully written.
This connection is established via the \V|Checkpoint::subCP()| method (as shown in 
Listing \ref{algo:nestedCp}), which defines a parent-child relationship.
After creating a parent checkpoint (CL1), all child checkpoints (CL2) are 
invalidated. 
In case CL2 data elements must be preserved for all successive
L2 iterations as well, these data elements can be made part of either CL1, or both
CL1 and CL2 checkpoints.

\begin{table}
\begin{center}
  \begin{tabular}{|C{1cm}|C{4cm}|}
        \hline
        Failure Stage & \parbox{20cm}{ Recovery from \TT{Checkpoint} \\ CL1-version/CL2-version}\\ \hline
        I & - / - \\ \hline 
        II & - / 10 \\ \hline 
        III & - / 20 \\ \hline 
        IV & 1 / - \\ \hline 
        V & 1 / 10 \\ \hline 
   \end{tabular} 
\caption{The versions of nested checkpoints that should be read depend 
on the program stage where the failure happened (corresponding to Fig. 
\ref{fig:nestedCpTimeline}). Reading the latest checkpoints of all 
nested levels can cause inconsistency, e.g., if failure happens at stage `IV',
the latest inner checkpoint should not be read.}
\label{tab:nestedCpRecoveryStages}
\end{center}
\end{table}

In case of asynchronous checkpointing with a combination of multi- or 
nested-level checkpoints, the user is responsible for 
the completion of previous/child checkpoints by calling \V|Checkpoint::wait()|.

In the SCR library, each node level checkpoint must be self contained, 
i.e., the checkpoint files cannot contain data that spans multiple checkpoints.
Due to this restriction, only one checkpoint in a nested 
checkpoint program can benefit from the SCR library, whereas the other
checkpoints must be taken at the PFS level.
Therefore, it is recommended to have high-frequency smaller checkpoints (e.g.,
inner iteration at CL2) at the node level, and low-frequency larger checkpoints 
at the PFS level (e.g., outer iteration at CL1).
In the example shown in Listing \ref{algo:nestedCp}, SCR 
is disabled for CL1 by \V|Checkpoint::disableSCR()| method.

Note that the \V|restartIfNeeded()| member function of the inner nested checkpoint 
(CL2 in the example) is called multiple times.
Still the checkpoint is only read at the first instance of a restarted run.
This method first checks the current CP-version of the
corresponding checkpoint, which is 0 only at the first or restarted run of the 
program. 
A non-zero value of CP-version implies a successful successive nested loop run, 
thus it returns immediately without reading the checkpoint.

\begin{figure}[htb]
  \centering
  \includegraphics[width=5.5cm,clip=true]{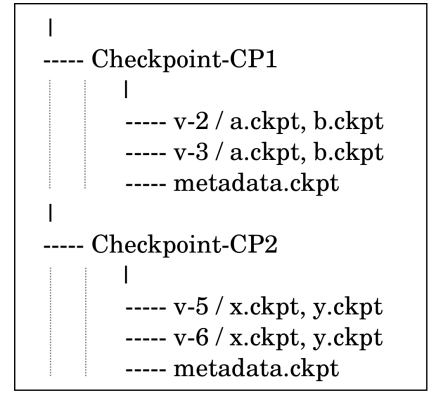}
	\caption[]{An example of the CRAFT directory structure with two-level checkpoints.
	\label{fig:cp-directory-st}}\vspace{.00cm}
\end{figure}

\subsection{Checkpoint file and directory structure}
All checkpoints of an application reside under a base checkpoint path, 
which can be set by the \V|CRAFT_CP_PATH| environment variable 
(see Table \ref{tab:craft-params}).
An example of the directory structure of CRAFT checkpoints is shown in 
Fig. \ref{fig:cp-directory-st}.
This example contains two checkpoint directories from two \V|Checkpoint|
objects containing different data, 
i.e. Checkpoint-CP1 contains data `a' and `b', whereas the Checkpoint-CP2
contains `x' and `y'.
The name of each \V|Checkpoint| object is given by the user at its definition,
whereas the names of each file are required while adding (using \V|add()|) 
the checkpointable objects.
Different CP-versions of a checkpoint are placed in separate version directories
(e.g., v-1, v-2 etc.). 
Each checkpoint maintains a meta-data file to keep track of the latest 
checkpoint version number, which is read at the beginning of a restart
phase.

\section{CRAFT(II): Automatic Fault Tolerance}
\label{sec:craftAFT}
\begin{table*}[t]
\begin{tabular}{@{}p{\dimexpr0.5\linewidth-\tabcolsep}
                  p{\dimexpr0.5\linewidth-\tabcolsep}@{}}
\begin{lstlisting}[style=base , caption=A sample application with basic 
CRAFT checkpoint/restart functionality., label=algo:sampleApp2]
#include <mpi.h>  
#include <craft.h>
int main(int argc, char* argv[]){
  ...
  int myrank, iteration = 0, cpFreq = 10;




	double data = 0;
	Checkpoint  myCP( "myCP", MPI_COMM_WORLD);
	myCP.add("data", &data);
	myCP.add("iteration", &iteration);
	myCP.commit(); 
  @myCP.restartIfNeeded(&iteration)@;
  for(; iteration <= n; iteration++){
    // Computation-communication loop
    myCP.updateAndWrite(iteration, cpFreq);  
	}
  ...


}
\end{lstlisting}
&
\begin{lstlisting}[style=base, caption=Same application as in Listing \ref{algo:sampleApp2} 
with CRAFT's automatic fault tolerance feature added., label=algo:sampleApp2WAFT]
#include <mpi.h>  
#include <craft.h>
int main(int argc, char* argv[]){
  ...
  int myrank, iteration = 0, cpFreq = 10;
  MPI_Comm FT_Comm;
	MPI_Comm_dup(MPI_COMM_WORLD, &FT_Comm);
  @AFT_BEGIN(FT_Comm, &myrank, argv);	@
|||*********************************************|||
	double data = 0;                          |||*|||
	Checkpoint  myCP( "myCP", FT_Comm);       |||*|||
	myCP.add("data", &data);                  |||*|||
	myCP.add("iteration", &iteration);        |||*|||
	myCP.commit();                         |||AFT Zone|||
  @myCP.restartIfNeeded(&iteration)@;         |||*|||
  for(; iteration <= n; iteration++){       |||*|||
    /* Computation-communication */         |||*|||
    myCP.updateAndWrite(iteration, cpFreq); |||*||| 
	}                                         |||*|||
  ...                                       |||*|||
|||*********************************************|||
	@AFT_END();@
}
\end{lstlisting}
\end{tabular}
\label{table:toycode-aft}
\end{table*}

This section provides the implementation details of the AFT interface,
which can be used for dynamic recovery of the application after process 
failure(s).
Listings \ref{algo:sampleApp2} and \ref{algo:sampleApp2WAFT}
draw a comparison between a simple 
toy-code and its corresponding AFT-enabled version.
The data checkpoints are taken in either case.
The AFT enabled code first defines a communicator, which is used for all 
further communication.
The most important additions in the code are the \texttt{AFT\_BEGIN()} and 
\texttt{AFT\_END()} macros. They define an `AFT zone'.
The \texttt{AFT\_BEGIN()} macro takes an MPI communicator and the input-argument
string as inputs and returns the rank of the current process in the given 
MPI communicator.
Within \texttt{AFT\_BEGIN()}, an error handler is attached to the provided 
MPI communicator. 
The code inside the AFT zone is wrapped in a \V|try-catch| block surrounded 
by a \V|while()| loop that iterates until the \V|try| block's execution completes 
successfully (i.e., without throwing an exception). 
An exception is thrown at a process failure.
In this case the communication repair mechanism is triggered in the \V|catch| 
block.
Since the MPI Standard does not provide a fault tolerant communication 
interface, the core requirement for the AFT feature is ULFM-MPI.

\subsection{ULFM-MPI}
Though in its prototype stage, ULFM-MPI is the most powerful candidate
for fault tolerant MPI functionality to be included in the MPI Standard.
Its prototype implementation has attracted significant attention in the MPI 
applications development community.
ULFM provides all helping functions to detect and recover from a process 
failure in an MPI application.

The most important ULFM functions are briefly described in the following:

\begin{itemize}[leftmargin=*] 
\item \TT{MPIX\_Comm\_revoke(MPI\_Comm)}: After process failure detection 
by any of the communicating partners, the revoke operation can be called on
the communicator by any of the detecting process(es) to invalidate the
communicator. 
The is a collective operation in the sense that it influences all processes 
within the associated communicator, but unlike any other collective, it 
does not require a symmetric call on all processes. 
A single process can initiate the revoking of the entire communicator. 

\item \TT{MPIX\_Comm\_shrink(MPI\_Comm, MPI\_Comm)}: This call takes a dirty 
communicator (i.e., containing failed processes) and returns a shrunk
healthy communicator by excluding the failed processes.
This is a collective operation and performs the consensus algorithm to ensure
the equal visibility of the eliminated processes throughout the communicator.

\item \TT{MPIX\_Comm\_agree(MPI\_Comm, int)}: This collective call can be used  
at any recovery stage to ensure that all processes within the communicator 
have reached that particular stage.

\end{itemize}

In addition to the above calls, new error codes are added, namely 
\V|MPIX_ERR_PROC_FAILED| and \V|MPIX_ERR_REVOKED|.
A detailed description of the complete ULFM-MPI API is available in 
\cite{bland2012proposal}.
The ULFM-MPI calls in combination with standard 
MPI functions such as \V|MPI_Comm_spawn()|, \V|MPI_Comm_get_parent()|, and 
\V|MPI_Intercomm_merge()| can be used to implement a communication 
recovery strategy. 

It should be noted that both the AFT-feature as well as its underlying ULFM-MPI 
library are in their prototype and experimental phases. 
Therefore some normal MPI operations may not work directly out of the box.
For example, RDMA and MPI File-IO are not currently supported under  
ULFM (release version 1.1). 
Thus, a special patch is used to suppress error notifications.
These patches are discussed and available at 
\cite{website:ulfm-fileio-modif, website:ulfm-RMA-modif}.
\subsection{Communicator recovery}
\label{sec:aftRecoverStrategy}
Using ULFM, the developer is flexible 
to choose the failure detection and recovery methodology. 
In \cite{Bland_post_failure}, Bland et. al. give an overview of 
application recovery strategies possible using ULFM.
Once a particular recovery method is chosen, the order in which ULFM's failure 
detection and communication recovery functions should be called is quite 
unequivocal. 
We have leveraged this fact and came up with a very simple interface
by abstracting away the details of all ULFM operations behind the
\texttt{AFT\_BEGIN()} and \texttt{AFT\_END()} macro. 
An error handler is used for detection of failures, 
whereas for communication recovery, a shrinking or non-shrinking 
option can be chosen. 
Here a `shrinking recovery' means that a new healthy 
communicator is created simply by removing the failed processes 
from the old communicator. 
On the other hand, a `non-shrinking recovery' rebuilds the communicator by
replacing failed process(es) by newly spawned ones.
For the non-shrinking recovery, there can be two strategies of
locality of the spawned processes. 
The environment variable \V|CRAFT_COMM_SPAWN_POLICY| can be either set
to \V|REUSE|, which spawns the recovery processes on the same node as before,
or to \V|NO-REUSE|, which spawns them on spare nodes, if available. 
The bookkeeping of reserve nodes (if available), failed nodes, as well as 
working nodes and processes is managed by CRAFT.
The user can specify the recovery model and spawned processes 
locality information via environment variable parameters 
(see Sec. \ref{sec:craft-params}).

\subsection{Data recovery}
After rebuilding the communication structure, the next step is to recover 
the data of the lost processes. 
This step depends on the application as well as the communication recovery 
method and can take the form of CR, ABFT, etc.

Within the scope of this work, we have chosen CRAFT's CR method 
(Sec. \ref{sec:craftCR}) in combination with the non-shrinking recovery model.
This combination eliminates the need to redistribute the domain 
(as in the case of shrinking recovery). 
Thus, the processes can make process-local checkpoints. 
This in turn enables the use of SCR for node-level 
checkpoints, which reduces the overhead.
Depending on the applications, other data-recovery strategies may be used 
in combination with AFT.

\section{CRAFT properties}
\label{sec:craftProperties}
This section describes the environment variables and the
memory management inside CRAFT.

\subsection{CRAFT parameters:}
\label{sec:craft-params}
In order to allow flexible modification of the CRAFT behavior, some of the CRAFT 
settings can be influenced by environment variables. 
Table \ref{tab:craft-params} shows the list of these parameters with their
brief description and default behavior. 
These variables are read from environment only once, 
either at the definition of a \V|Checkpoint| object or at the start of the AFT zone.
Thus, changing their values during an application run do not have any influence.

\begin{table}[t]
\begin{center}
  \begin{tabular}{|C{2.5cm}|C{5cm}|}
        \hline
        Parameter Name & Description  \\ \hline 
        {\TT{CRAFT\_CP\_PATH}} & The base path to all checkpoints. \newline Default: The working directory, i.e., \TT{\$PWD}. \\ \hline 
        {\TT{CRAFT\_ENABLE}} & Enables/disables CRAFT checkpointing. \newline Values: 1(default),0 \\ \hline 
        {\TT{CRAFT\_WRITE\_ ASYNC}} & Enables asynchronous writing of checkpoints. \newline \scriptsize{Values: 0(default), 1} \\ \hline
        {\TT{CRAFT\_WRITE\_ ASYNC\_ZERO\_ COPY}} & Enables asynchronous writing without creating an additional data copy. \newline \scriptsize{Values: 0(default), 1} \\ \hline
        {\TT{CRAFT\_ASYNC\_ THREAD\_PIN\_ CPULIST}} & Asynchronous threads' locality (pinning) information. \newline \scriptsize{format: e.g. 10\_20, the asynchronous threads of two MPI processes on each node will be pinned to the 10th and 20th core of the node. No pinning by default} \\ \hline
        {\TT{CRAFT\_USE\_SCR}} & Controls the usage of SCR. If CRAFT is compiled with SCR, it can still be deactivated by this variable. \newline Values: 1(default), 0 \\ \hline 
        {\TT{CRAFT\_READ\_CP\_ ON\_RESTART}} & Whether the restarted run should resume by reading checkpoints or not.\newline Values: 1(default), 0 \\ \hline 
        {\TT{CRAFT\_COMM\_ RECOVERY\_POLICY}} & Specifies the method of communicator recovery. \newline \scriptsize{Values: \V|NON-SHRINKING|(default), \V|SHRINKING|} \\ \hline 
        {\TT{CRAFT\_COMM\_ SPAWN\_POLICY}} & Determines the node-locality of spawned processes in case of Non-shrinking recovery. \newline \scriptsize{Values: \V|NO-REUSE|(default), \V|REUSE|} \\ \hline
    \end{tabular} 
\\
\caption{The CRAFT environment variables. Note: 1=enable, 0=disable}
\label{tab:craft-params}
\end{center}
\end{table}

\subsection{Checkpoint memory management}
\label{sec:cp-mem-management}
As explained in Sec. \ref{fig:craftDesignLogic}, the \V|cpMap| collects the 
shared pointers (\V|std::shared_ptr|) of all checkpointable objects.
In this way, the scope of the checkpointable objects is automatically 
managed. 
For example, if the user creates a checkpointable object inside the application
(using any of the CRAFT extension methods in Sec. \ref{sec:craft-extension}),  
and adds it to a particular checkpoint, 
the checkpointable object is not destroyed at the end of the 
checkpoint object scope. 
In this way, the user can use the checkpointable object further in the 
program as well. 
The user can either delete such shared pointer objects manually
or they get deleted automatically once they go out of scope.

With AFT, however, the proper memory 
management is a challenging task and requires careful attention.
On the part of the user, the objects declared inside the AFT zone must 
either be defined statically or via smart pointers (for dynamic allocation).
This way the objects get destroyed once they are out of scope.
As of the memory allocated inside MPI calls, MPI or ULFM-MPI currently do not 
offer any mechanism to cleanup the possible communication memory stacks in case 
a communicator is suddenly revoked. 
We address this issue as part of our future work.

\section{Experimental Framework}
\label{sec:experimental_framework}

We have performed all benchmarks on the Emmy 
cluster\footnote{{Emmy} cluster at the {E}rlangen {R}egional {C}omputing 
{C}enter ({RRZE}): \url{https://www.anleitungen.rrze.fau.de/hpc/emmy-cluster/}} 
at RRZE. 
Equipped with 560 compute nodes, each having two Xeon 2660v2 ``Ivy Bride'' 
chips (10 cores per chip + SMT) running at 2.2\,GHz and 64\,GB-RAM, the Emmy cluster 
has the overall peak performance of 234\,TFlop/s. 
The system has Infiniband interconnect with 40\,Gbits/s bandwidth per link and 
direction.
The parallel file-system (LXFS) has a capacity of 400\,TB and an 
aggregated parallel IO bandwidth of more than 7000\,MB/s.

\subsection{Benchmark application} 
The complete functionality of CRAFT is showcased using a Lanczos solver.
The Lanczos algorithm is an iterative method for finding some eigenvalues 
of a sparse matrix. 
We use it to find the minimum eigenvalues of a test matrix. 
The pseudo-code of the Lanczos algorithm is shown in Algorithm 
\ref{pseudocode:lanczos-algo}. 
Each iteration calculates the new Lanczos vectors, $\alpha$,
and $\beta$. 
The approximated minimum eigenvalues are then calculated using 
the QL method and checked against the convergence criterion. 
In order to have a deterministic runtime, we fix the number of iterations
in our benchmarks.
However, in a practical code, a residual test with a desired tolerance of 
calculated eigenvalues is used as the convergence criterion to
abort the iteration loop.
The checkpoint data mainly consists of the Lanczos vectors,
$\alpha$, and $\beta$ values etc.
For this purpose, CRAFT is extended for the vector data types of 
the utilized numerical linear algebra library GHOST \cite{Kreutzer2016}.

\begin{algorithm}
\begin{algorithmic}
\For{j:=1,2, ..., numIter}
                \Function {lanczos-step}{}
                        \State ${\omega}_j \gets A{\nu}_j$
                        \State ${\alpha}_j \gets {\omega}_j . {\nu}_j$
                        \State ${\omega}_j \gets {\omega}_j - {\alpha}_j{\nu}_j - {\beta}_j{\nu}_{j-1}$
                        \State ${\beta}_{j+1} \gets \|{\omega}_j\|$
                        \State ${\nu}_{j+1} \gets {\omega}_j / {\beta}_{j+1} $
                \EndFunction

                ${CalcMinimumEigenVal()}$
\EndFor
\end{algorithmic}
\caption{The pseudo-code of the Lanczos algorithm for finding eigenvalues of a matrix $A$. 
}\label{pseudocode:lanczos-algo}
\end{algorithm}

The sparse matrix for our benchmarks arises from the quantum-mechanical
description of the electron transport properties of graphene.
Graphene is the blueprint for quasi 2D materials with many distinctive 
characteristics and has prospective application areas in nanotechnology 
and nanoelectronics. 
In our benchmark, the matrix is generated on the fly using a library tool 
rather than reading from a file. 
This saves the expensive step of reading the matrix from the PFS.

\subsection{MPI library}
For all AFT-benchmarks, the ULFM-MPI release 1.1 version is used. 
The ULFM-MPI extensions features are built on top of Open MPI version 1.7.1.
As described in \ref{sec:cr-optimizations}, the asynchronous checkpointing 
optimizations require the MPI library to support 
\V|MPI_THREAD_MULTIPLE|. 
The Open MPI version 1.7.1 is recognized to have erroneous behavior under
\V|MPI_THREAD_MULTIPLE| \cite{website:openmpibug1}.
Therefore, in order to benchmark the CR optimizations 
(in Sec. \ref{sec:cr-overhead-benchmark}), Intel MPI version 5.1 is used.

\subsection{Fault model} 
In this paper, we focus on fail-stop failures, i.e., failures 
that cause a process to fail permanently. 
These processes become nonresponsive to any communication request, thus 
they can be detected during the following communication request 
involving the failed process by the ULFM failure detection mechanism.
We usually simulate the failure of a complete node crash by killing all processes 
on a particular node via \verb|pkill -9 <program>|. 
However, in a few benchmarks (in order to have a deterministic re-computation 
overhead), we have injected the ``process failures'' at a predetermined iteration 
from inside the program.

\section{Results and Evaluation}
\label{sec:results}
This section shows performance results for simple benchmarks
and the application scenario described above.
In a fault tolerance library/tool, it is important to 
explicitly show the overhead faced by the application due 
to the presence of FT capabilities. 
Within the scope of this work, these overheads can be categorized into 
the following groups.
\begin{itemize}[leftmargin=1\parindent]
\item Checkpoint and restart overhead\,(OH$_{cp}$, OH$_{res}$): 
An application faces checkpoint overhead even in the absence of failures. 
A failure recovery imposes additional restart overhead. 
The restart overhead includes the data re-initialization as well as reading
the data from the checkpoint.
Optimization methods such as asynchronous and node-level 
checkpointing aim at reducing this overhead.
\item Communication recovery overhead\,(OH$_{rec}$): 
This overhead comprises only the time it takes to recover the 
broken communicator. 
The overhead varies depending on the communication recovery policy.
\item Recomputation overhead\,(OH$_{redo}$): This overhead depends on the 
point in time between two successive checkpoints when the program faces a 
failure.
It can be influenced by choosing an appropriate checkpointing
interval. In addition, ABFT techniques can be used to accelerate the 
recovery process \cite{MarkusHuber_Superman2016}.
\end{itemize}

\subsection{Scaling behavior of communication recovery overhead}
\label{sec:comm-rec-time}
First, we analyze the scaling behavior of communication 
recovery (i.e., AFT) overhead (OH$_{rec}$), without any influence from the 
highly application-dependent checkpointing overhead. 
A simple benchmark is used for this purpose, in which an 
\V|MPI_Barrier| is repeatedly called in a loop inside an AFT zone.
Once process(es) fail, all other processes of the 
communicator are notified and follow recovery routines as
explained in Sec. \ref{sec:aftRecoverStrategy}.

\begin{figure}[tb]
	\centering
	\includegraphics[width=8.3cm,clip=true]{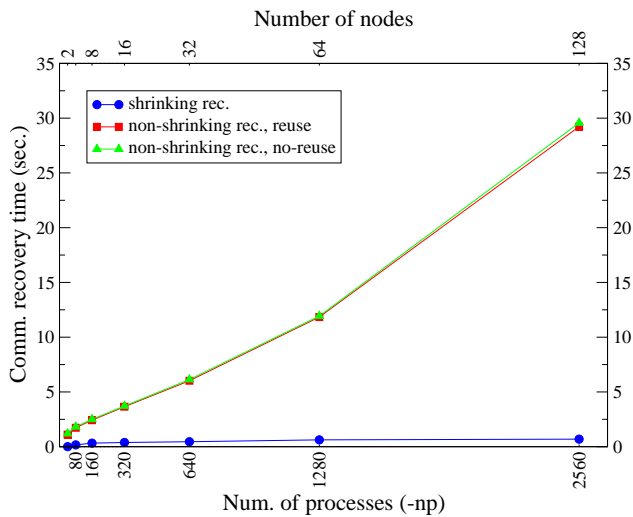}
	\caption[]{The scaling of MPI communication recovery time w.r.t. 
            the number of processes (20 processes per node) for Shrinking, 
            Non-Shrinking with reuse of failed nodes,  
            and Non-Shrinking with no-reuse of failed nodes. 
	\label{chart:nodeScaling}}\vspace{.00cm}
\end{figure}

\begin{table}
\begin{center}
  \begin{tabular}{|C{5cm}|C{2cm}|}
        \hline
        Description & Time(sec.)  \\ \hline 
        Communicator revoke + shrink & 0.34 \\ \hline 
        Generate processes-spawn info. & 0.23 \\ \hline 
        Spawn + merge & 26.10 \\ \hline 
        Redistribute proc. ranks & 1.39 \\ \hline 
        Resource management & 0.68 \\ \hline 
    \end{tabular} 
\\
\caption{Breakdown of the communication recovery phase with 'non-shrinking, no-reuse' 
recovery policy with 2560 processes on 128 nodes.}
\label{tab:rec-OH-breakdown}
\end{center}
\end{table}

Figure \ref{chart:nodeScaling} shows the scaling 
of the communication recovery overhead on up to 2560 processes 
(with 20 processes per node) using different recovery modes. 
The shrinking recovery mode shows the best scaling behaviour with very
little influence from the number of processes involved. 
The recovery time is $\approx0.51$ seconds for a total of 2560 processes
running on 128 nodes.
The non-shrinking recovery overhead, however, increases linearly 
with respect to the number of processes.
Table \ref{tab:rec-OH-breakdown} shows the recovery overhead breakdown.
The major part of it comes from the blocking operations of 
\V|MPI_Comm_spawn()| and \V|MPI_Intercomm_merge()|.
The rest of the overhead comes from creating spawn information 
(via \V|MPI_Info_set()|), reassigning spawned processes
the same identity as failed processes, and managing the available 
recovery nodes.
The node `reuse' policy for spawned processes is slightly 
cheaper than the `no-reuse' policy.
This is due to extra steps involved to manage the record of available recovery 
nodes. 

\begin{figure}[tb]
	\centering
	\includegraphics[width=8.3cm,clip=true]{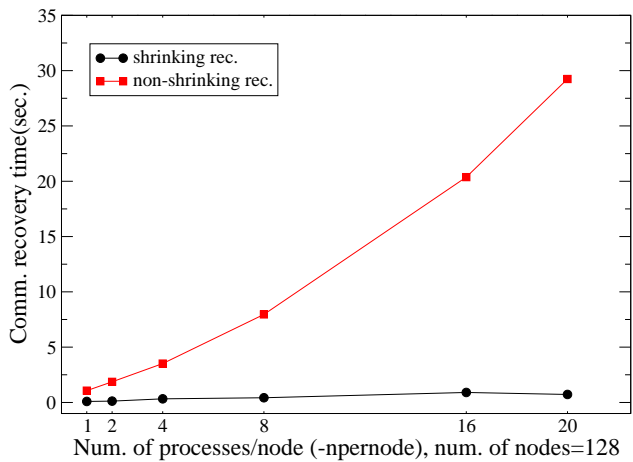}
	\caption[]{The scaling of MPI communication recovery time w.r.t the number of 
              processes per node. A total of 128 nodes were used.
	\label{chart:npernodeScaling}}\vspace{.00cm}
\end{figure}

Figure \ref{chart:npernodeScaling} shows a comparison of communication recovery 
overhead with respect to the number of processes per node. 
A total of 128 nodes are used in this case, varying the number of processes per node.
This benchmark shows that the recovery overheads are much more influenced by the 
number of processes involved than the number of nodes. 
Thus, the communication recovery overhead can be significantly lowered by having 
fewer processes per node and using shared-memory parallelization models 
(e.g. OpenMP) within the nodes.

\begin{figure}[tb]
	\centering
	\includegraphics[width=8.3cm,clip=true]{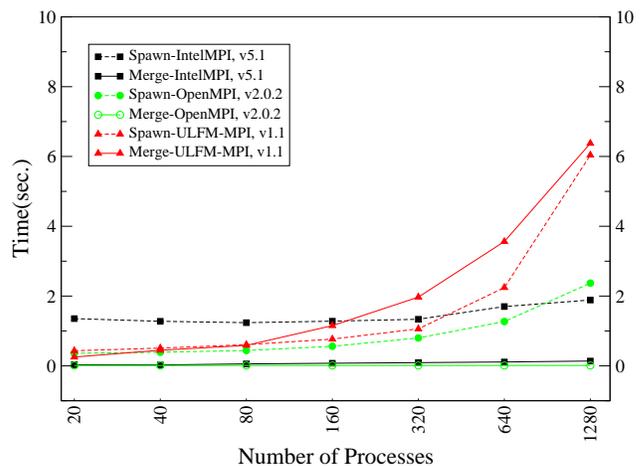}
	\caption[]{A scaling comparison of spawn and merge routines for 
              Intel MPI v5.1 vs. OMPI-v1.10.3 vs. ULFM-1.1 implementations.
	\label{chart:spawnMergeTimes}}\vspace{.00cm}
\end{figure}

As shown above, a large part of the non-shrinking recovery time is contributed by 
the MPI spawn and merge routines that exist in the MPI Standard since MPI-2.0
(rather than the recent test-phase ULFM extension), so we conducted a scaling 
benchmark for different MPI implementations.
It spawns 20 processes each time and merges them into 
the original communicator. 
Figure \ref{chart:spawnMergeTimes} shows a scalability comparison of spawn and merge 
routines for three different implementations and reveals that both Intel MPI  
(version 5.1) and Open MPI (version 1.10.3) have a good scaling behavior. 
The ULFM-MPI (version 1.1), however, which is forked from an earlier Open MPI 
branch (version 1.7.1) shows poor scaling of these functions.
It is therefore reasonable to expect that the upcoming versions of ULFM-MPI, which 
will be ported to the up-to-date Open MPI branch, will show better scalability.

Once a node has a hard failure, it is likely to encounter failures again. 
Thus, it is advisable not to use the same node again whenever possible.
In the following, we only use the `no-reuse' policy of the non-shrinking recovery.

\subsection{Checkpoint/restart benchmark}
\label{sec:cr-overhead-benchmark}
We employ Lanczos application to determine the impact of 
CR-overhead (OH$_{cp}$) using the optimization techniques described in Sec. 
\ref{sec:cr-optimizations}.
This benchmark is run with the following parameters: 
num. of iterations=3000, checkpoint frequency=500, 
matrix num. of rows \& columns = $9.0 \cdot 10^8$, 
number of non-zero elements = $11.7 \cdot 10^9$, 
global checkpoint size $\approx 14.4\,$GB.
Table \ref{tab:CP-overhead} shows the runtime and the overhead incurred 
by three different checkpointing methods on 128 nodes (2560 processes). 
Compared to the baseline runtime case (no 
checkpointing), the synchronous checkpointing adds 1.3\% of overhead.
Asynchronous checkpointing reduces this overhead to 0.56\%, 
whereas node level checkpointing reduces the overhead further to only 0.20\%.

\begin{table}
\begin{center}
  \begin{tabular}{|C{1.6cm}|C{1.1cm}|C{1.1cm}|C{1.1cm}|C{1.1cm}|}
        \hline
                    & No CP & Sync. CP (PFS)  & Async. CP (PFS) & Node-level CP (SCR) \\ \hline 
        Runtime     & 788.9 & 799.4           & 793             & 790.48              \\ \hline 
        \% overhead & -     & 1.33            & 0.56            & 0.20                \\ \hline
        average time/CP (sec.)& -     & 2.0            & -            & 0.18                \\ \hline 
   \end{tabular} 
\\
\caption{An overhead comparison of three different checkpointing, namely,
synchronous, asynchronous and node-level checkpointing (128 nodes of Emmy with 
20 processes per node, Intel MPI).}
\label{tab:CP-overhead}
\end{center}
\end{table}

\subsection{Automatic Fault Tolerant application}
\label{sec:ft-application}
We now make use of both the CR and AFT parts of CRAFT to showcase
their use with the Lanczos (Sec. \ref{sec:experimental_framework}) application.
We use the same application parameters as in Sec. \ref{sec:cr-overhead-benchmark}.
The failures are introduced from inside the application 
at the mid-point of two successive checkpoints.

\begin{figure}[tb]
	\centering
	\includegraphics[width=9.4cm,clip=true]{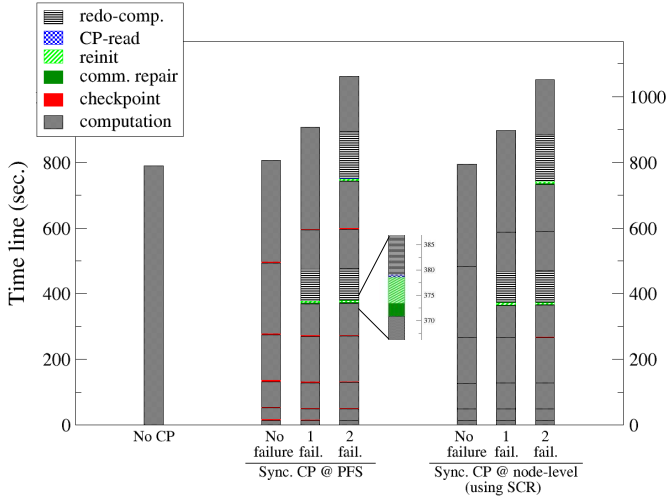}
	\caption[]{Various checkpoint and failure-recovery scenarios of the Lanczos 
            benchmark using 128 Emmy nodes (256 processes). 
            The average communication recovery time in process failure cases is 2.6 seconds.
	\label{chart:lanczosGhost}}\vspace{.00cm}
\end{figure}

Figure \ref{chart:lanczosGhost} shows the runtime of the Lanczos benchmark on 128 nodes
(256 Processes, 10threads/process) in various cases.
The first case, `No CP', shows the runtime without any checkpoints
and failures.
This is the baseline for all other cases.
Depending on the location of checkpoint storage (PFS, node level), the benchmarks 
are grouped into two categories.
The first group `Sync. CP @ PFS' shows the program timeline with 
checkpoints at PFS. 
Each PFS checkpoint introduces an overhead of around $2$ sec., compounding a 
total overhead (OH$_{cp}$) of $\approx\,1.26$\%.
The second group of bars `Sync. CP @ node-level' shows the timeline with neighbor-level 
checkpoints using SCR.
Here each checkpoint causes an overhead of $\approx\,0.9$ seconds.
The main overhead appears in failure recovery cases (2nd and 3rd bar of 
each group), where the major part of the overhead is spent on the 
re-computation of lost work (OH$_{redo}$).
Independent of the checkpoint location, the communication repair (OH$_{rec}$) 
takes $\approx\,2.6$ seconds.

Note that these benchmark results are presented to showcase the behavior of CRAFT 
in a particular (albeit typical) application setting on a particular machine. 
As such, they cannot be generalized to other scenarios without actually re-running 
the tests.

\subsection{Additional CR extensions}
In addition to the standard types, the CRAFT library was endowed with support 
for GHOST sparse matrix data types 
\cite{Kreutzer2016}, Phist sparse matrix data types \cite{jonas_phist_dlr}, 
and Intel MKL complex data types. 
These extensions are part of the downloadable code \cite{website:craftlib}.

\section{Related Work}
\label{sec:relatedWork}

The research and development of CR services can be categorized based 
on the level of transparency and software layer implementation.

A few libraries such as Open MPI \cite{hursey-phd-thesis-2010, website:ompi-cr} and 
LAM/MPI \cite{LAMMPI} (using BLCR \cite{website:blcr}) and 
MPICH-V \cite{MPICH-V-Bosilca} (using Condor \cite{condor-techrep})
provide a very transparent way to make system-level checkpoints. 
Libckpt \cite{libckpt} was a CR tool that offered a hybrid between 
system and user-level checkpointing.
Its optimizations included incremental and forked checkpointing.
The functionality to exclude certain parts of memory made 
it possible to minimize the overall checkpoint data volume.

With a strong focus on portability, tools like $PC^3$ \cite{PCCube}
(based on $C^3$ \cite{c-cube}), CPPC \cite{CPPC-2010} and 
Porch \cite{Porch-CP} support semi-automatic ALCR services via 
pre-compiler and/or code analysis approaches.
The user must provide directives at the potential 
checkpoint locations, which gets instrumented by the pre-compiler 
during source-to-source compilation stage.
Unlike the classic ALCR approaches, 
the compiler-assisted ALCR services are less transparent,
and save the entire state of the program that creates 
checkpoints almost as large as system-level checkpoints.
CPPC also offers the explicit definition 
of data via variable registration for the application-level checkpoints.
These variables are registered as memory segments of basic data types.
The checkpoints are then written in a portable (platform-independent) format.
The pre-compiler assists the user in registering the restart-relevant variables 
based on a liveness analysis and also modifies the flow-control of 
the application.

The AFT \cite{AFT-poster-trung} is a relatively new tool that 
semi-automatically inserts CR functionality into the code based
on code analysis and user input.
Another Open MPI \cite{website:openmpi} ALCR service 
is called SELF \cite{website:ompi-cr}, which uses callback functions 
to activate checkpoints. 
The actual writing of the data is the responsibility of the user.
At a desired stage, the user can initiate 
the checkpoint from outside the program, which starts Open MPI's
coordinated CR-service followed by the calls to the user's implemented 
functions.

Our library is partly similar to the OpenMPI-SELF CR service in the way 
that it requires the user to provide I/O routines in case of an arbitrary
data type. 
However, CRAFT provides out-of-the-box I/O methods for most basic  
data types. 
Once a new data type is made CRAFT-checkpointable, it can avail
the underlying optimization techniques like asynchronous and node-level 
checkpointing.

Several efforts revolved around fault tolerance on the MPI communication level. 
FT-MPI \cite{FTMPI:Fagg2000} (later known as HARNESS \cite{HARNESS:Fagg2001}) 
was the first widely known activity in this regard. 
FT-MPI offered three different failure recovery modes: SHRINK,
BLANK, and REBUILD. 
The fault detection and communication recovery was part of FT-MPI, which 
minimized the code changes in the application.
However, the project is no longer maintained.
MPI/FT \cite{MPIFT:Batchu2004} was a similar effort which provided 
process failure recovery for two application models: 
master-worker and Single Program Multiple Data (SPMD).
In case a worker process failed in the master-worker approach, 
the MPI library would notify the master, which would relaunch the worker.
In the SPMD applications, it repaired the \V|MPI_COMM_WORLD| by replacing the 
failed processes and reading the checkpoint. 
However, it required the SPMD applications to have synchronous loops 
on all processes. 
ULFM is the latest and most serious effort to include fault tolerance at the 
MPI level.
It has gathered significant attention in the community and many scientific 
application developers have benefited from its prototype implementation
\cite{exampi14, fault-tolerant-montecarlo, 
application-level-fault-recovery, Nuria_Losada_ULFM_17, 
towards-local-failure-local-recovery-using-ULFM}.

In the AFT-feature of CRAFT, we aim to provide a solution similar to FT-MPI, based 
on ULFM. We hide away the communication recovery details from the user
and offer a simpler interface that requires minimal code changes 
and is applicable to a large variety of applications.


\section{Summary}
\label{sec:summary}
CRAFT is a library that provides two important building blocks for creating 
a fault tolerant application. 
Its checkpoint/restart\,(CR) part is an extendable library base, 
using which the application-level checkpoint/restart functionality can be 
easily introduced in the programs with minimal modifications.
The most frequently used data types (POD, POD arrays, multi-arrays, MPI derived 
datatypes, etc.) are 
part of CRAFT by default and can be used out of the box. 
Furthermore, user defined data types can easily be made CRAFT-checkpointable.
The asynchronous checkpointing feature or the Scalable Checkpoint/Restart\,(SCR)
can be used to reduce the checkpointing overhead.

Based on ULFM-MPI, the automatic fault tolerance\, (AFT) part of the library 
provides an easier interface for the dynamic process recovery in case of process 
failures.
CRAFT hides many details of the process failure detection and 
communication recovery process, and enables the user to select via an
environment variable whether to perform a shrinking or a non-shrinking 
recovery of the failed communicator.
The current AFT implementation includes support for Torque and
SLURM job managers. 

In this work, we have described the implementation details of CRAFT and 
showcased its capabilities via examples. 
We have also shown how its CR functionality can be extended for a user defined
data type. 
The scaling of the AFT feature was benchmarked on up to 2560 processes on 128 nodes. 
Though the newly added extended functions of ULFM-MPI behave well at scale,
the scaling behavior of \V|MPI_Comm_spawn()| and \V|MPI_Intercomm_merge()| 
routines is unsatisfactory.
A Lanczos eigenvalue solver was used to showcase the usage of CRAFT 
in real-world applications. 
Node-level CR via SCR significantly reduces the CR-overhead. 
The benchmarks revealed that the communication recovery overhead itself remained 
in an acceptable range.
The major part of the total overhead comes from the re-computation of the lost work,
which depends on the point in time between two successive checkpoints when the 
failure occurs. 

The CRAFT library is open source under a BSD license and is available at 
\cite{website:craftlib}.
Though CRAFT's CR feature is well tested and production proof, its AFT feature (built upon 
ULFM-MPI) is in its prototype phase and must be used with caution. In certain application
cases (e.g., DMA and MPI IO operations) the ULFM-MPI requires additional modifications.

\section{Acknowledgments}
This work is supported by the German Research
Foundation\,(DFG) through the Priority Program 1648 ``Software for
Exascale Computing"\,(SPPEXA) under project ESSEX-II (Equipping
Sparse Solvers for Exascale) \cite{website:essex}. 
Special thanks goes to Dr. George Bosilca and Dr. Klaus Iglberger 
for valuable suggestions and input which helped us overcome design 
and implementation challenges.


\bibliographystyle{IEEEtran}
\bibliography{IEEEabrv,./CR_refs}

\vspace{-10 mm}
\begin{IEEEbiographynophoto}{Faisal Shahzad}
is a PhD student in the HPC group at Erlangen
Regional Computing Center (RRZE). He received his B.Sc. degree
in Mechanical Engineering in 2006 from GIK 
Institute of Science and Technology, Topi, Pakistan and 
M.Sc. degree in Computational Engineering in 2011 from University of 
Erlangen-Nuremberg, Germany. His research area focuses to 
research various fault tolerance techniques for algorithms
in computational science.
\end{IEEEbiographynophoto}
\vspace{-12 mm}
\begin{IEEEbiographynophoto}{Jonas Thies}
is a scientific employee at the German Aerospace Center, Institute of Simulation 
and Software Technology, Department of High Performance Computing.
He received a PhD in mathematics from the University of Groningen, 
the Netherlands, in 2010, a Masters Degree in Scientific Computing from KTH 
Stockholm in 2006 and a Bachelors Degree in Computational Engineering from 
the University of Erlangen-Nuremberg in 2003. 
His research interests include sparse matrix computations, HPC software 
engineering and CFD.
\end{IEEEbiographynophoto}
\vspace{-12 mm}
\begin{IEEEbiographynophoto}{Moritz Kreutzer}
completed his B.Sc. in Computational Engineering in 2009 and M.Sc. in Computational 
Engineering in 2011 from the University of Erlangen-Nuremberg, Germany.
Currently, he is working as a PhD student in the HPC group at RRZE.
\end{IEEEbiographynophoto}
\vspace{-12 mm}
\begin{IEEEbiographynophoto}{Thomas Zeiser}
holds a PhD in Computational Fluid Mechanics from the
University of Erlangen-Nuremberg. He is now a senior research
scientist in the HPC group of RRZE and is among many other things
still interested in lattice Boltzmann methods.
\end{IEEEbiographynophoto}
\vspace{-12 mm}
\begin{IEEEbiographynophoto}{Georg Hager}
holds a PhD in Computational Physics from
the University of Greifswald. He has been working with
high performance systems since 1995, and is now a senior
research scientist in the HPC group at RRZE. Recent research includes
architecture-specific optimization for current microprocessors,
performance modeling on processor and system levels,
and the efficient use of hybrid parallel systems. His
daily work encompasses all aspects of user support in HPC
such as lectures, tutorials, training, code parallelization,
profiling and optimization, and the assessment of novel
computer architectures and tools.
\end{IEEEbiographynophoto}
\vspace{-12 mm}
\begin{IEEEbiographynophoto}{Gerhard Wellein}
holds a PhD in Solid State Physics from
the University of Bayreuth and is a regular Professor at the
Department for Computer Science at University of Erlangen.
He heads the HPC group at RRZE and has more than 10 years of
experience in teaching HPC techniques to students and
scientists from Computational Science and Engineering.
His research interests include solving large sparse eigenvalue
problems, novel parallelization approaches, performance
modeling, and architecture-specific optimization.
\end{IEEEbiographynophoto}

\end{document}